\documentclass[aps,prl,reprint,groupedaddress,amsmath,amssymb,showpacs]{revtex4-1}

\usepackage{graphicx}
\usepackage{natbib}

\begin{document}

\title{Tuning Energy Relaxation along Quantum Hall Channels}

\author{C. Altimiras}
\affiliation{CNRS, Laboratoire de Photonique et de Nanostructures
(LPN) - Phynano team, route de Nozay, 91460 Marcoussis, France}
\author{H. le Sueur}
\affiliation{CNRS, Laboratoire de Photonique et de Nanostructures
(LPN) - Phynano team, route de Nozay, 91460 Marcoussis, France}
\author{U. Gennser}
\affiliation{CNRS, Laboratoire de Photonique et de Nanostructures
(LPN) - Phynano team, route de Nozay, 91460 Marcoussis, France}
\author{A. Cavanna}
\affiliation{CNRS, Laboratoire de Photonique et de Nanostructures
(LPN) - Phynano team, route de Nozay, 91460 Marcoussis, France}
\author{D. Mailly}
\affiliation{CNRS, Laboratoire de Photonique et de Nanostructures
(LPN) - Phynano team, route de Nozay, 91460 Marcoussis, France}
\author{F. Pierre}
\email[Corresponding author: ]{frederic.pierre@lpn.cnrs.fr}
\affiliation{CNRS, Laboratoire de Photonique et de Nanostructures
(LPN) - Phynano team, route de Nozay, 91460 Marcoussis, France}

\date{\today}

\begin{abstract}
The chiral edge channels in the quantum Hall regime are considered ideal ballistic quantum channels, and have quantum information processing potentialities. Here, we demonstrate experimentally, at filling factor $\nu_L=2$, the efficient tuning of the energy relaxation that limits quantum coherence and permits the return toward equilibrium. Energy relaxation along an edge channel is controllably enhanced by increasing its transmission toward a floating ohmic contact, in quantitative agreement with predictions. Moreover, by forming a closed inner edge channel loop, we freeze energy exchanges in the outer channel. This result also elucidates the inelastic mechanisms at work at $\nu_L=2$, informing us in particular that those within the outer edge channel are negligible.
\end{abstract}


\maketitle
Nanocircuits at low temperature exhibit new phenomena resulting from the quantum nature of the transport. In mesoscopic physics, a major objective is to make use of these phenomena to develop novel quantum functionalities for the future nanoelectronics. Quantum information processing, which has the potential to outperform classical computing \cite{[See e.g. ]bennet2000qip}, is a striking illustration. A key task to reach this objective is to find efficient ways to enhance the robustness of quantum effects. It is also necessary to increase controllably the relaxation rate toward equilibrium, e.g. to perform fast resets. Here, we demonstrate experimentally the up-down control of energy relaxation along a single quantum channel realized in the integer quantum Hall regime (QHR).

In the QHR, electrons confined to two dimensions and immersed in a strong perpendicular magnetic field propagate in chiral channels along the sample edges \cite{halperin1982qhc}. The electrical current circulates without dissipation along these edge channels (EC), generally considered as ideal ballistic quantum channels \cite{buttiker1988abs}. The similitude between ECs and light beams has inspired novel electronic devices \cite{ji2003emz,samuelsson2004twoeAB,*olkhovskaya2008hom,neder2007i2e}, and proposals for quantum information processing \cite{bertoni2000qlgwetiqw,*stace2004mowc,*Ionicioiu2001qcwbe}. However, recent experiments found that quantum interferences with edge states are more fragile than anticipated \cite{ji2003emz,litvin2007dsec,bieri2009fbv,neder2007i2e,roulleau2008lphi}. This observation, corroborated by the recent observation of strong energy exchanges between copropagating ECs \cite{lesueur2010relaxiqhr}, and possibly with other thermalized states at filling factor $\nu_L=1$ \cite{granger2009och}, seemingly impedes the potentialities of these ECs for quantum electronics. However, in the present work we show how to tune efficiently the energy relaxation and thereby the dephasing, with regards to both increasing and freezing the energy exchanges along an EC at $\nu_L=2$, where two copropagating ECs are present.

\begin{figure}[!t]
\includegraphics[width=\columnwidth]{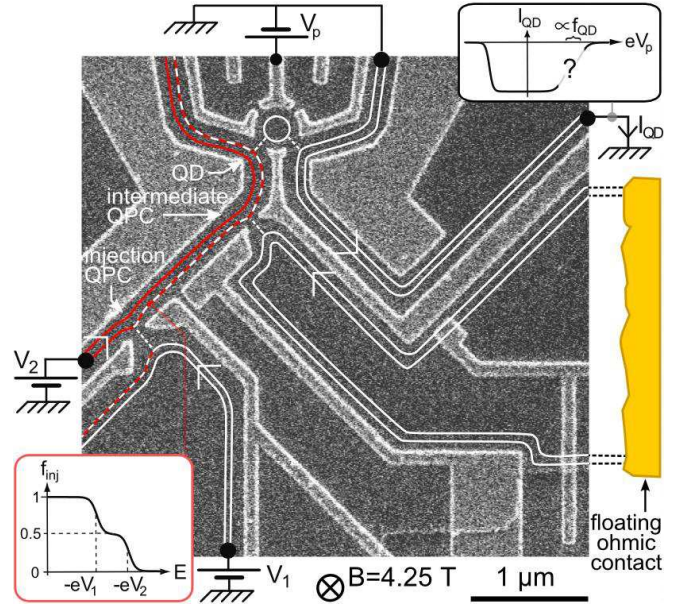}
\caption{(Color online) Sample e-beam micrograph: metallic gates appear bright; the wide gates on the left and right of the quantum dot (QD) are grounded and can be ignored. Electronic excitations propagate counter clockwise along two edge channels (EC), depicted by lines. Dashed lines connecting ECs indicate transmission through quantum point contacts (QPC). At the output of the injection QPC, the energy distribution $f_\mathrm{inj}$ is a double step (left inset) in the half transmitted outer EC. Electronic excitations travel along adjustable paths from the injection QPC to the QD, and part of the outer EC can be diverted toward a floating ohmic contact by tuning the intermediate QPC's transmission. Right inset: the tunnel current $I_{\mathrm{QD}}$ through the QD is proportional to the energy distribution $f_\mathrm{QD}$ probed at the QD in the left outer EC.}
\label{Fig1}
\end{figure}

The basic experimental principle to probe energy relaxation is sketched in Fig.~1 and detailed in the previous works \cite{altimiras2010nesiqhr,lesueur2010relaxiqhr}. It consists in driving the outer EC out-of-equilibrium with a voltage biased `injection' quantum point contact (QPC) and, after various propagation paths defined mostly by voltage biased metal gates, in measuring at the path's end the electronic energy distribution $f_\mathrm{QD}(E)$ in the outer EC with a quantum dot (QD) operated as an energy filter. At the output of the injection QPC, whose conductance is set to $0.5e^2/h$ such that the outer EC is half transmitted and the inner EC fully reflected, the electronic energy distributions in the outer and inner ECs are, respectively, a non-equilibrium smeared double step $f_\mathrm{inj}$ (see left inset in Fig.~1) and a cold Fermi function \cite{altimiras2010nesiqhr}. Energy exchanges along the edge are revealed through changes in $f_\mathrm{QD}(E)$ with the length of the propagation path between injection QPC and QD. In the actual measurements, we record the differential conductance $\partial I_{\mathrm{QD}}/\partial V_p \propto \partial f_\mathrm{QD}/\partial E$, with $V_p \propto E$ the voltage applied to a plunger gate coupled to the QD, and $I_{\mathrm{QD}}$ the tunnel current across the QD. In the following, we display these raw data normalized by the measured maximum QD current $I_{\mathrm{QD}}^\mathrm{max}$. Full details regarding the $f_\mathrm{QD}$ spectroscopy are given in \cite{altimiras2010nesiqhr,lesueur2010relaxiqhr}. See \cite{si} for specific additional information on QD calibration.

The sample shown in Fig.~1 was patterned by e-beam lithography on a standard GaAs/Ga(Al)As two dimensional electron gas of density $2~10^{15}~\mathrm{m}^{-2}$ and mobility $250~\mathrm{m}^2V^{-1}s^{-1}$. The same sample was used to demonstrate non-equilibrium EC spectroscopy \cite{altimiras2010nesiqhr} and energy relaxation \cite{lesueur2010relaxiqhr} in the QHR. Conductance measurements were performed using standard low frequency lock-in techniques in a dilution refrigerator of base temperature $30~\mathrm{mK}$ and near the center of the $\nu_L=2$ plateau. To avoid artificial heating, AC voltages were always kept smaller than $k_BT/e$.

\begin{figure}[!t]
\includegraphics[width=0.85\columnwidth]{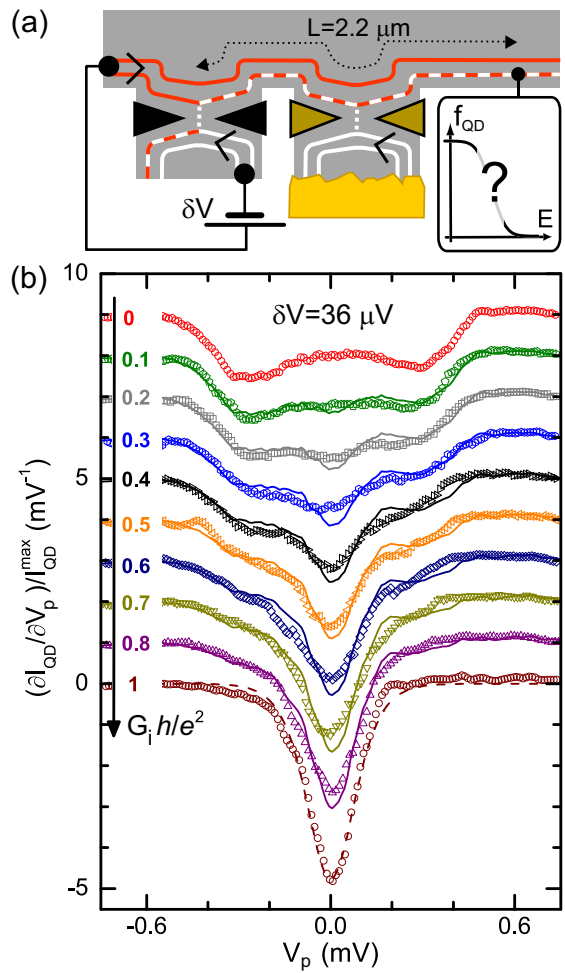}
\centering
\caption{(Color online) (a) Experiment schematic: the outer EC's relaxation is forced by partly diverting it toward a floating ohmic contact, through the intermediate QPC (middle split gate) of conductance $G_\mathrm{i}\in [0,1]~e^2/h$. (b) Raw data $\propto \partial f_\mathrm{QD}(E)/\partial E$ (symbols) at $\delta V=36~\mu$V are plotted vs $V_p \propto E$ and shifted vertically for different values of $G_\mathrm{i}$. The dashed line is a Fermi fit ($T_\mathrm{fit}=36$~mK). Continuous lines are the weighted sums of the data at $G_\mathrm{i}=0$ and $e^2/h$, as predicted by the scattering theory.}
\label{Fig2}
\end{figure}

\textbf{Energy relaxation with a voltage probe.} We show that we can controllably drive the system up to full relaxation by diverting the non-equilibrium EC toward a floating ohmic contact. The ohmic contact here plays the role of the so-called `voltage probe' introduced by theorists to account for decoherence and energy relaxation within the scattering approach to quantum transport \cite{buttiker1986rqcsr}. Voltage probes act as reservoirs that absorb all incoming electronic quasiparticles and emit new ones with a Fermi statistics at the electrochemical potential dictated by current conservation. These absorption/emission processes mimic both the quasiparticles' finite quantum lifetime and their relaxation toward thermal equilibrium along an EC. An `intermediate' QPC of tunable conductance $G_\mathrm{i}$ permits to connect controllably the ohmic contact (see Fig.~1 and setup schematic in Fig.~2(a)). The impact of similar floating ohmic contacts has previously been investigated on current fluctuations \cite{oberholzer2006pcc}. More recently, their dephasing properties were explored using an electronic Mach-Zehnder interferometer \cite{roulleau2009tdwvp}. Here, we use an experimental setup that permits us to characterize fully the corresponding energy relaxation.

This experiment is performed as follows: the injection QPC is located $L=2.2~\mu$m upstream of the QD, as shown in Fig.~1, and biased at $\delta V=V_1-V_2=36~\mu$V. The resulting non-equilibrium outer EC propagates for $1.4~\mu$m along the edge before it reaches the intermediate QPC. It is there partly transmitted with a probability $G_\mathrm{i}h/e^2\in[0,1]$ toward a floating ohmic contact connected to the bottom right ECs shown in Fig.~1. The energy distribution in the outer EC is then measured at the QD, $0.8~\mu$m downstream of the intermediate QPC.

Figure~2(b) shows as symbols the raw data $(\partial I_{\mathrm{QD}}/\partial V_p)/I_{\mathrm{QD}}^\mathrm{max}$ vs $V_p$ obtained for different $G_\mathrm{i}$ spanning from zero to $e^2/h$. At $G_\mathrm{i}=0$, we observe a double dip that corresponds to a significant but incomplete energy redistribution (as in \cite{lesueur2010relaxiqhr}). In contrast to this non-equilibrium double dip, the energy derivative of a Fermi function is a single dip whose width and inverse amplitude are proportional to the temperature. In the opposite limit, $G_\mathrm{i}=e^2/h$, we observe a narrow single dip fitted using a cold Fermi function at $T_\mathrm{fit}=36$~mK (dashed line). This shows that ECs emitted by the floating ohmic contact are fully thermalized, and that they are not heated up along their way back to the intermediate QPC. At partial transmissions $0<G_\mathrm{i}h/e^2<1$, the data in Fig.~2(b) exhibit a more complex shape, which we now compare to the scattering model's predictions.

According to the scattering theory, the energy distribution at the intermediate QPC's output is the sum of the distribution functions in the two incoming ECs weighted by their respective transmission probabilities \cite{buttiker1986rqcsr}. Because interactions are ignored in the scattering theory, it can be applied to the present experiment only if the energy relaxation is negligible along the $0.8~\mu$m path between the intermediate QPC and the QD. We showed in our previous work on the same sample and in the same experimental conditions that this is a good approximation \cite{altimiras2010nesiqhr}. However, the energy relaxation is significant after a propagation path of $2.2~\mu$m \cite{lesueur2010relaxiqhr}. This is taken into account by using the data measured at $G_\mathrm{i}=0$ as the reference signal for the outer EC coming from the injection QPC. The corresponding predictions are the sum of the data at $G_\mathrm{i}=0$ and $e^2/h$ weighted, respectively, by the measured $1-G_\mathrm{i}h/e^2$ and $G_\mathrm{i}h/e^2$ (continuous lines in Fig.~2(b)). We find a good agreement with the data, \textit{without any fitting parameter}. In conclusion, we show here that the connection through a QPC of a floating ohmic contact used as a voltage probe provides a mean to drive controllably a non-equilibrium quantum channel back toward equilibrium, in quantitative agreement with the scattering theory.

\begin{figure}[!t]
\includegraphics[width=0.9\columnwidth]{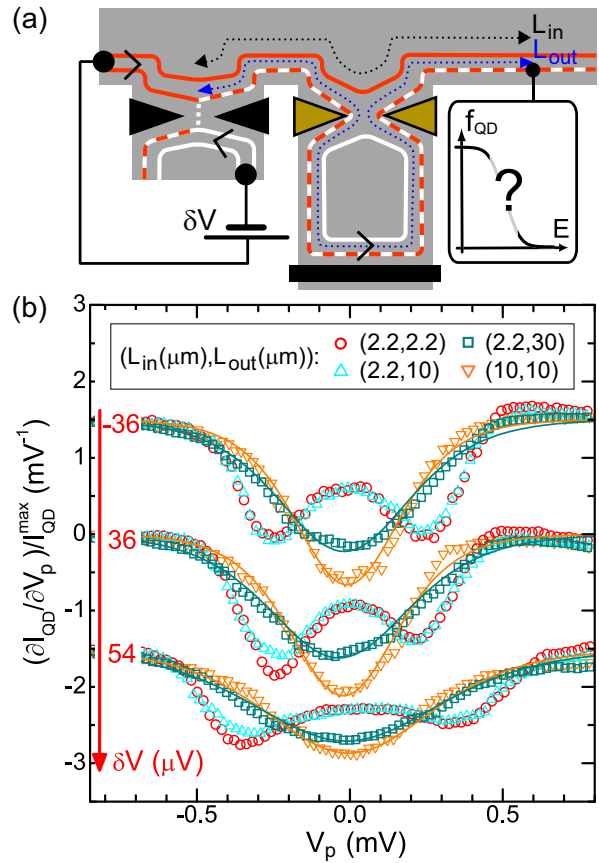}
\centering
\caption{(Color online) (a) The non-equilibrium outer EC propagates partly along a closed inner EC's loop, except when $L_{in}=L_{out}$. (b) Raw data $\propto \partial f_\mathrm{QD}/\partial E$ (symbols) for various $(L_{in}~(\mu\mathrm{m}),L_{out}~(\mu\mathrm{m}))$ are shifted vertically for the different $\delta V\in\{-36,36,54\}~\mu$V. The data for $(2.2,10)$ are mostly unchanged from those for the short direct path $(2.2,2.2)$. This implies that \textit{energy exchanges are negligible} in the outer EC along the $8~\mu$m closed inner EC loop. The contrast is stark with the data for the corresponding direct path $(10,10)$, which exhibit a broad Fermi dip (continuous lines are fits with hot Fermi functions). The observed energy relaxation toward a broad dip for the larger $28~\mu$m inner EC loop of $(2.2,30)$ shows the above energy exchanges freezing depends on loop size.}
\label{Fig3}
\end{figure}

\textbf{Freezing energy exchanges.} Energy exchanges remain important even with a single EC \cite{granger2009och} (possibly due to the presence of low energy spin excitations in the bulk at $\nu_L=1$ \cite{plochocka2009spinpolarnu1}). However, we show here that energy exchanges can be frozen at $\nu_L=2$ by closing the inner EC on itself along the outer EC's path. The experiment schematic is shown in Fig.~3(a). Contrary to the previous setup, the ECs transmitted through the intermediate QPC do not reach the floating ohmic contact. Instead, they are redirected toward the QD by applying a negative voltage to a surface metal gate barring the way (either the bottom right gate in Fig.~1 or another gate further away \cite{si}). For $G_\mathrm{i}=e^2/h$, the outer EC is fully transmitted and the inner EC fully reflected. As a result, the inner EC propagates from injection QPC to QD on a distance $L_{in}\simeq2.2~\mu$m, shorter than the outer EC's $L_{out}\simeq10$ or $30~\mu$m: the extra outer EC propagation path takes place along a closed inner EC loop of perimeter $L_{loop}\simeq 8~\mu$m or $28~\mu$m. For $G_\mathrm{i}=0$ and $2e^2/h$, both the inner and outer ECs copropagate along the same length $(L_{in}=L_{out})$ of 2.2 and 10~$\mu$m, respectively. In the following, we characterize the experimental configuration by the two propagation lengths $(L_{in}~(\mu\mathrm{m}),L_{out}~(\mu\mathrm{m}))$.

Figure 3(b) shows as symbols $(\partial I_{\mathrm{QD}}/\partial V_p)/I_{\mathrm{QD}}^\mathrm{max}$ measured vs $V_p$ for $\delta V=-36$, $36$ and $54~\mu$V applied to the injection QPC. The striking feature is that the data for $(2.2,2.2)$ and those for $(2.2,10)$ are essentially identical, at our experimental resolution. This demonstrates directly that the energy distribution in the probed outer EC remains unchanged along the extra $8~\mu$m path, and therefore that energy exchanges are negligible. The contrast is stark compared to $(10,10)$, where the outer EC propagates on the same length as for $(2.2,10)$, but with a copropagative inner EC. Indeed, the non-equilibrium double dip structures apparent for $(2.2,2.2)$ and $(2.2,10)$ are now washed out for $(10,10)$, and the energy distributions have relaxed toward hot Fermi functions (continuous lines in Fig.~3(b) are Fermi fits). Consequently, we find that closing the inner EC on a $8~\mu$m loop turns down energy exchanges along the outer EC, from a complete energy redistribution (toward a hot electron equilibrium) to a negligible one! A different behavior arises with the larger $28~\mu$m inner EC loop realized for $(2.2,30)$, where $f_\mathrm{QD}(E)$ relaxes toward a hot Fermi function with a temperature higher than that for $(10,10)$. This revival of energy exchanges shows loop size has an important role.

We attribute the observed reduction of energy exchanges to the discreteness of energy levels in the closed inner EC loop, whose spacing competes with the available energy. Indeed, for a $8~\mu$m closed loop and using the standard drift velocity $v_D\approx10^5$~m/s at $\nu_L=2$ \cite{talyanskii1993vd}, the energy spacing $\delta E_{in}\approx52~\mu e$V is larger than or comparable to $e|\delta V|$. Consequently, the available energy in the outer EC is not sufficient to excite the discrete inner EC's energy levels \footnote{Similar conclusions can be reached beyond a first order perturbative treatment of interactions (P.\ Degiovanni, private communication).}. As a result, the energy exchanges between ECs that were previously established at $\nu_L=2$ \cite{lesueur2010relaxiqhr} are here turned off. This analysis is confirmed by the revival of energy exchanges for a $28~\mu$m loop: the corresponding spacing $\delta E_{in}\approx15~\mu e$V is smaller than $e|\delta V|$, therefore there is enough energy available in the outer EC to excite the discrete inner EC's energy levels.

The observation of frozen energy exchanges also provides new information regarding the inelastic mechanisms along an EC at filling factor $\nu_L=2$. First, this shows that on a $8~\mu$m length scale at the probed energies, interactions between electrons within the outer EC are incontrovertibly negligible. The same conclusion applies to many other plausible energy exchange mechanisms, including the coupling with the nearby surface metal gates, the nuclear spins, the disorder induced bulk states enclosed in the loop, the nearby counter propagating ECs and the additional modes within the outer EC \cite{aleiner1994nee,*chamon1994er} that are predicted in presence of edge reconstruction for realistic smooth confinement potentials \cite{chklovskii1992eec,*macdonald1993er}. Second, this sheds light on the extra energy leak observed in $(10,10)$ and $(30,30)$ \cite{lesueur2010relaxiqhr} compared to expectations for two interacting ECs \cite{degiovanni_plasmon_2010,*lunde_interaction_2010}. This leak strongly suggests significant energy transfers toward extra degrees of freedom. Following our previous works \cite{altimiras2010nesiqhr,lesueur2010relaxiqhr}, we extracted from $f_\mathrm{QD}(E)$ the increase of energy with $\delta V$. We find that the electronic energy is identical at our experimental accuracy in the configurations $(2.2,2.2)$, $(2.2,10)$ and $(2.2,30)$. This shows that the invoked extra degrees of freedom are frozen by closing the inner EC into a loop of up to $30~\mu$m \footnote{Note that energy exchanges between ECs along the loop of $(2.2,30)$ do not change $T_\mathrm{exc}$ in the stationary regime, if there is no energy leak from the inner EC's loop.}. Consequently, the predicted additional modes within the outer EC can be ruled out (although not those within the inner EC). The temperatures $T_\mathrm{exc}$, corresponding to the energy increase of the probed non-equilibrium excitations, are shown with error bars in \cite{si}. We illustrate the above finding with $\delta V=36~\mu$V, for which we obtain approximately the same $T_\mathrm{exc}=89$, $91$ and $93$~mK for, respectively, $(2.2,2.2)$, $(2.2,10)$, $(2.2,30)$, whereas we find $T_\mathrm{exc}=73$~mK for $(10,10)$.

Perspectives of the present technique to reduce drastically the energy exchange rate include the increase of the electronic phase coherence length well beyond the largest values reported in mesoscopic circuits. Indeed, although the dephasing is known to be driven not only by energy exchange mechanisms but also by the significant contribution of the low frequency current noise of the second quantum Hall channel at $\nu_L=2$ \cite{roulleau2008noisedephasing,*neder2007cdngsn,*levkivskyi_noise-induced_2009}, closing the inner EC in loops forbids current fluctuations and thereby cancels out this additional contribution. In practice, inner EC closed loops of several microns are easily implemented, compatible even with optical lithography. From the decrease of the energy relaxation rate by more than a factor four \footnote{With two copropagating ECs the energy relaxation is found significant for a propagation path of $2.2~\mu$m, whereas it is found negligible along a $8~\mu$m closed inner EC loop. Consequently, the energy exchange rate is here decreased by more than a factor four.}, we anticipate an increase of the quantum coherence length from $20~\mu$m \cite{roulleau2008lphi} to near macroscopic length scales, above $80~\mu$m.

\begin{acknowledgments}
The authors gratefully acknowledge discussions with A.\ Anthore, P.\ Degiovanni, P.\ Joyez, F.\ Portier, P.\ Roche.
\end{acknowledgments}

%\bibliography{biblio}

%Merlin.mbs v4.21 2009-07-09.
%

\end{document}

\title{Supplementary Material for \\`Tuning Energy Relaxation along Quantum Hall Channels'}

\author{C. Altimiras}
\thanks{These authors contributed equally to this work.}
\affiliation{CNRS, Laboratoire de Photonique et de Nanostructures
(LPN) - Phynano team, route de Nozay, 91460 Marcoussis, France}
\author{H. le Sueur}
\thanks{These authors contributed equally to this work.}
\affiliation{CNRS, Laboratoire de Photonique et de Nanostructures
(LPN) - Phynano team, route de Nozay, 91460 Marcoussis, France}
\author{U. Gennser}
\affiliation{CNRS, Laboratoire de Photonique et de Nanostructures
(LPN) - Phynano team, route de Nozay, 91460 Marcoussis, France}
\author{A. Cavanna}
\affiliation{CNRS, Laboratoire de Photonique et de Nanostructures
(LPN) - Phynano team, route de Nozay, 91460 Marcoussis, France}
\author{D. Mailly}
\affiliation{CNRS, Laboratoire de Photonique et de Nanostructures
(LPN) - Phynano team, route de Nozay, 91460 Marcoussis, France}
\author{F. Pierre}
\email[Corresponding author: ]{frederic.pierre@lpn.cnrs.fr}
\affiliation{CNRS, Laboratoire de Photonique et de Nanostructures
(LPN) - Phynano team, route de Nozay, 91460 Marcoussis, France}

\maketitle

\section{Energy distribution spectroscopy with a quantum dot}

We measured the energy distribution $f_\mathrm{QD}$ in the outer EC using a QD as a tunable energy filter \cite{altimiras2010nesiqhr}. Assuming a single active QD level of energy $E_{lev}$, constant tunneling density of states and tunnel rates, the QD current $I_{\mathrm{QD}}$ in the sequential tunnel regime reads \cite{kouwenhoven1997etqd}
\begin{equation}
I_{\mathrm{QD}}=I_{\mathrm{QD}}^{max}(f_S(E_{lev})-f_\mathrm{QD}(E_{lev})), \label{SIIdot}
\end{equation}
where $I_{\mathrm{QD}}^{max}$ is the maximum QD current and $f_\mathrm{QD}$ ($f_S$) refers to the energy distribution in the outer EC located at the left (right) side of the QD. $f_\mathrm{QD}$ and $f_S$ can be extracted separately by applying a sufficiently large source-drain voltage, as shown in the right inset of Fig.~1.

The probed energy $E_{lev}=E_0-e\eta_G V_G$ was swept using $V_G$, with $E_0$ an unimportant offset and $\eta_G$ the gate voltage-to-energy lever arm calibrated by temperature and non-linear QD characterizations. Full details regarding the procedure to extract $\eta_G$ are given in the supplementary material of \cite{lesueur2010relaxiqhr}. The precise value of $\eta_G$ is not of utmost importance in the present work, yet it enters as a scaling factor for the overall energy, including the fit and excess temperatures (e.g. in Supplementary Figure~3). For completeness, we recapitulate in Table~\ref{SItabEtaG} the extracted values of $\eta_G$ for the different configurations with their relative standard errors.

As pointed out above, the simple expression of Supplementary Equation~\ref{SIIdot} assumes only one QD level contributes to $I_{\mathrm{QD}}$, and neglects the energy dependence of the electrodes tunneling density of states and of the tunnel rates in and out the QD. In practice, the validity of these hypotheses were checked with a standard non-linear QD characterization \cite{kouwenhoven1997etqd}, and by comparing mixing chamber temperatures with fit temperatures obtained within this framework (see Figure~2 in \cite{altimiras2010nesiqhr}).

Importantly, the measured energy distribution is that of 1DCF quasiparticle excitations in the probed edge channel. In particular, the tunnel coupled QD does not probe the predicted additional edge excitations \cite{aleiner1994nee} corresponding to transverse charge oscillations across the finite width of edge channels \cite{chklovskii1992eec} (see \cite{altimiras2010nesiqhr} and references therein for a detailed discussion).

\begin{table}
\center
\begin{tabular}{|c|c|c|c|c|c|}
  \hline
$(L_{in},L_{out})$ & $\eta_G$ & $\Delta \eta_G/\eta_G$  \\
$(\mu\mathrm{m})$ &   & $(\%)$  \\
\hline
$(2.2,2.2)$ & 0.0610 & 3.5 \\
$(2.2,10)^+$ & 0.0603 & 5.0 \\
$(2.2,30)$ & 0.0603 & 5.0 \\
$(10,10)^+$ & 0.0606 & 5.7 \\
\hline
\end{tabular}
\caption{Used lever arms $\eta_G$ and their relative standard errors $\Delta \eta_G/\eta_G$ for the freezing of energy exchanges (Fig.~3 and Supplementary Figure~3). The symbol ($^+$) points out experimental configurations, characterized by $(L_{in},L_{out})$, for which the full QD calibration was performed. A full calibration was performed only each time a surface gate nearby the QD was changed by an important amount. Note that the data for energy relaxation with a voltage probe (Fig.~2) were obtained in a different cooldown. The lever arm of this renewed QD was found to be $\eta_G \simeq 0.059$.}
\label{SItabEtaG}
\end{table}

\section{Practical realization of the different experimental configurations}

The experimental configuration was tuned in-situ with the bias voltages applied to surface metallic gates.

Supplementary Figure~1 displays an e-beam micrograph of the sample with a lower magnification than in Fig.~1. It shows the bottom right surface metal gate that permits to realize the $30~\mu$m propagation paths between injection QPC and QD.

Supplementary Figure~2 shows how we implement the configuration $(L_\mathrm{in}=2.2~\mu\mathrm{m},L_\mathrm{out}=10~\mu\mathrm{m})$.

\begin{figure}[!ht]
\center
\includegraphics[width=\columnwidth]{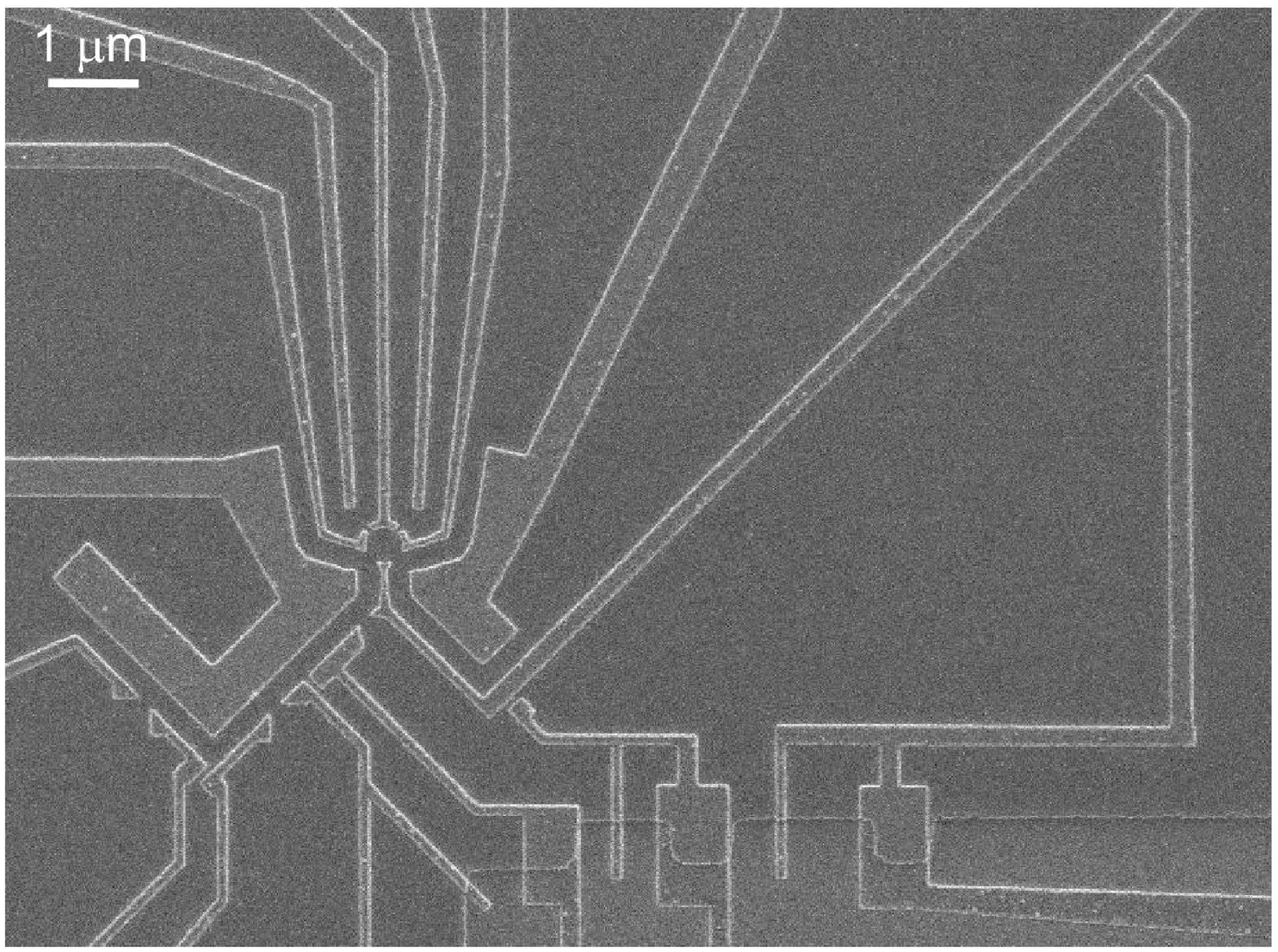}
\caption{E-beam micrograph of the sample.}
\label{SIfig1}
\end{figure}

\begin{figure}[!ht]
\includegraphics[width=\columnwidth]{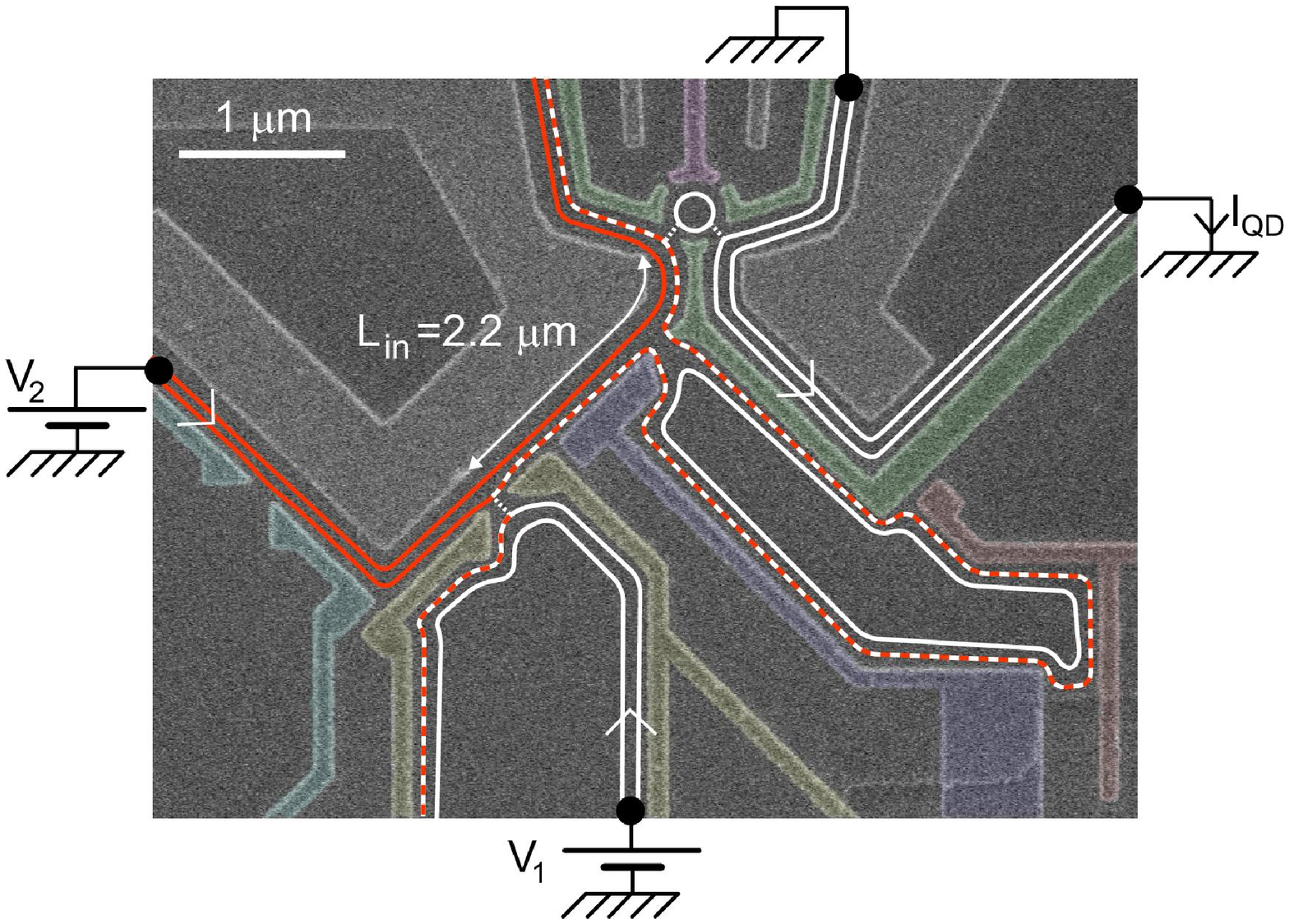}
\caption{Experimental realization of $L_{in}=2.2~\mu$m and $L_{out}=10~\mu$m. The corresponding schematic is shown in Fig.~2(a) and the corresponding data in Fig.~2(b).}
\label{SIfig2}
\end{figure}

\section{Extracted excess temperatures with a closed inner EC loop}

\begin{figure}[!th]
\includegraphics[width=\columnwidth]{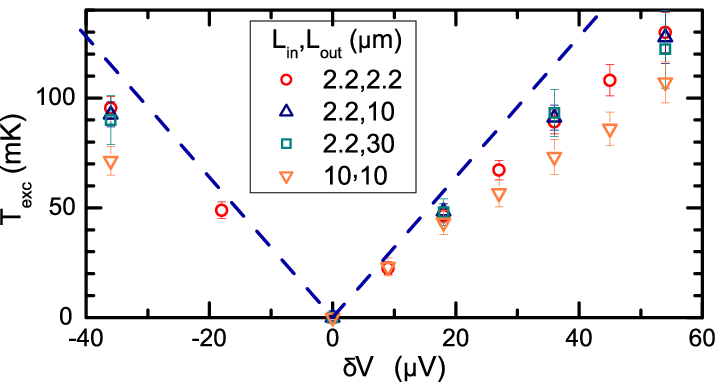}
\caption{Excess temperatures extracted from the raw data for various $(L_\mathrm{in},L_\mathrm{out})$ configurations. Fig.~3(b) in the manuscript shows some of the corresponding raw data.}
\label{SIfig3}
\end{figure}

The amount of energy within the probed electronic excitations can be extracted from the raw data. This information can be used to shed new lights on the energy exchange mechanisms at work.

Following our former works \cite{altimiras2010nesiqhr,lesueur2010relaxiqhr}, we express this energy $E_\mathrm{EC}$ as an excess temperature defined as:
\begin{equation}
T_\mathrm{exc}\equiv \sqrt{6 (E_\mathrm{EC}-E_\mathrm{EC}(\delta V=0))/\nu\pi^2k_B^2},
\end{equation}
with the ratio $E_\mathrm{EC}/\nu$ related to the energy distribution $f_\mathrm{QD}$ through:
\begin{equation}
E_\mathrm{EC}/\nu= \int (E-\mu)(f_\mathrm{QD}(E)-\theta(\mu-E))dE,\label{EsNu}
\end{equation}
with $\theta(E)$ the step function, $\nu$ the density of states per unit length and energy, and $\mu$ the electrochemical potential (see supplementary material in \cite{lesueur2010relaxiqhr} for full details on the procedure to extract $T_\mathrm{exc}$).

In other words, the excess temperature is the temperature difference that gives, at equilibrium, the same increase of energy as that extracted from the measured non-equilibrium distribution functions.

Supplementary Figure~3 shows as symbols the excess temperatures from the data with a closed inner EC as well as in the corresponding configurations with a copropagative inner EC (see Fig.~3 for some of the corresponding raw data).  For completeness, we also give in Table~\ref{SItabTfit} the fit temperatures corresponding to the continuous lines shown in Fig.~3(b).

\begin{table}
\center
\begin{tabular}{|c|c|c|c|c|c|}
  \hline
$(L_{in},L_{out})$ & $\delta V$ & $T_\mathrm{fit}$  \\
$(\mu\mathrm{m})$ & $(\mu\mathrm{V})$ & $($mK$)$  \\
\hline
$(10,10)$ & $-36$ & 81 \\
$(10,10)$ & $36$ & 83 \\
$(10,10)$ & $54$ & 129 \\
$(2.2,30)$ & $-36$ & 101 \\
$(2.2,30)$ & $36$ & 108 \\
$(2.2,30)$ & $54$ & 145 \\
\hline
\end{tabular}
\caption{Fit temperatures corresponding to the continuous lines shown in Fig.~3(b).}
\label{SItabTfit}
\end{table}

The interesting finding here is that the excess temperature is identical, at our experimental accuracy, in the configurations $(2.2,2.2)$ and $(2.2,30)$ although the latter shows important energy exchanges. On the contrary, the configuration $(10,10)$ displays substantially smaller excess temperatures. This implies that the energy exchanges that took place along the extra $28~\mu$m, in presence of a large closed inner EC loop, conserve the energy within the probed outer EC. This was expected in the stationary regime for energy exchanges in the outer EC mediated by the closed inner EC loop. An important consequence of this result is that there are negligible energy leaks toward additional states e.g. in the bulk or along the edge in this configuration. In particular, the present analysis permits us to extend to $28~\mu$m propagation lengths the conclusion that the energy exchanges with the predicted additional modes \cite{aleiner1994nee,chamon1994er} in the outer EC are negligible (see the paragraph before the conclusion of the manuscript for $8~\mu$m propagation lengths).

%\bibliography{biblio}

%Merlin.mbs v4.21 2009-07-09.
%